\documentclass[a4paper]{jpconf}
\usepackage{graphicx}
\usepackage{amsmath,amssymb,latexsym}
\newcommand \hmu {\hat{\mu}}
\begin{document}
\title{Strangeness at high temperatures}

\author{Christian Schmidt (for BNL-Bielefeld Collaboration)}

\address{Universit\"at Bielefeld, Fakult\"at f\"ur Physik, Postfach 100131, D-33501 Bielefeld, Germany}

\ead{schmidt@physik.uni-bielefeld.de}

\begin{abstract}
We use up to fourth order cumulants of net strangeness fluctuations and their correlations with net baryon number fluctuations to extract information
on the strange meson and baryon contribution to the low temperature hadron resonance gas, 
the dissolution of strange hadronic states in the crossover region of the QCD transition and the quasi-particle nature of strange quark contributions to the high 
temperature quark-gluon plasma phase.
\end{abstract}

\section{Introduction}
At RHIC and the LHC, quark gluon plasma is created by means of heavy ion collisions. 
The initial conditions of such collisions are free of net strangeness. 
It is theoretically well understood \cite{hic-rev} and experimentally verified \cite{strange-expt} 
that many strange, 
anti-strange quarks are created during these collisions. This will ultimately lead to event-by-event 
fluctuations of strangeness,
as long as only a fixed subsystem (specific kinetic window) is observed.
To what extent the created fireball reaches chemical equilibrium before its hadronic 
freeze out --- depending on collision energy and system size --- is, however, 
still under debate. Furthermore, it is questionable whether strange hadrons also freeze-out in 
the critical crossover region $(T\approx T_c)$ that is controlled by the light quark sector. 
Based on the very smooth behavior of strangeness fluctuations \cite{WHRG,hotQCDHRG} 
it has been suggested \cite{WB-Tc, Ratti} that deconfinement for strange quarks
takes place at a temperature significantly larger than $T_c$, determined by chiral symmetry breaking. 
This would imply the existence 
of strange bound sates inside the QGP for some temperatures $T> T_c$. A 
detailed analysis of various cumulants of net strangeness fluctuations and baryon number strangeness correlations obtained by 
the BNL-Bielefeld Collaboration \cite{our} supports, however, the statement that the strangeness liberation 
temperature is consistent with $T_c$. We will briefly repeat this study here. 
We obtain these cumulants  as derivatives 
of the logarithm of the QCD partition function, or equivalently of the pressure (P) in units 
of $T^4$, with respect to the chemical potentials. In general we define
\begin{equation}
\chi_{mn}^{XY} = \left. \frac{\partial^{(m+n)} [P(T,\hmu_X,\hmu_Y)/T^4]} 
{\partial \hmu_X^m \partial \hmu_Y^n} \right|_{\vec{\mu}=0}\ ,
\label{eq:chi}
\end{equation}
where $X,Y=B,S,Q$ are the quantum numbers of net baryon ($B$), net strangeness ($S$) and net electric charge ($Q$) and $\hmu_X=\mu_X/T$ 
are the corresponding chemical potentials with 
$\vec{\mu}=(\mu_B,\mu_S,\mu_Q)$. We will also use the notations $\chi_{0n}^{XY}\equiv\chi_n^Y$ and $\chi_{m0}^{XY}
\equiv\chi_m^X$. The various $\chi_{mn}^{XY}$ cary information on the quantum numbers of the relevant degrees of freedom in the 
system \cite{Koch,Ejiri}. Note, however, that the quantities $\chi_{mn}^{XY}$ also define the Taylor expansion coefficients of the pressure with 
respect to $\hmu_X$ and $
\hmu_Y$. Their study is thus important also for other reasons such as a critical point search \cite{GG,CS} or the analysis of the freeze out properties of heavy ion 
collisions \cite{freeze-out}. In the following, we will compare our non-perturbative lattice QCD results to an uncorrelated gas of hadrons at low temperatures and to an 
uncorrelated gas of quarks at high temperatures, with special emphasis on the strangeness fluctuations. Here the lattice calculations have been performed with highly 
improved staggered quarks (HISQ) and almost physical quark masses (the ratio of light to strange quark mass is 1/20, with a strange quark mass that was fixed to its 
physical value). The lattice sizes have been chosen as $N_\sigma^3\times N_\tau$, with aspect ratio $N_\sigma/N_\tau=4$ and $N_\sigma=24$ and $32$. A 
continuum limit is still to be taken and will require at least one even finer lattice spacing. However, from the stability of our current results we estimate very small effects of the nonzero lattice spacing. 
For the comparison with the HRG we take all particles and resonances up to a mass cutoff of 2.5 GeV into account.
  
\section{Strangeness in the hadronic phase}
\begin{figure}[t]
\begin{center}
\begin{minipage}{14pc}
\includegraphics[width=14pc]{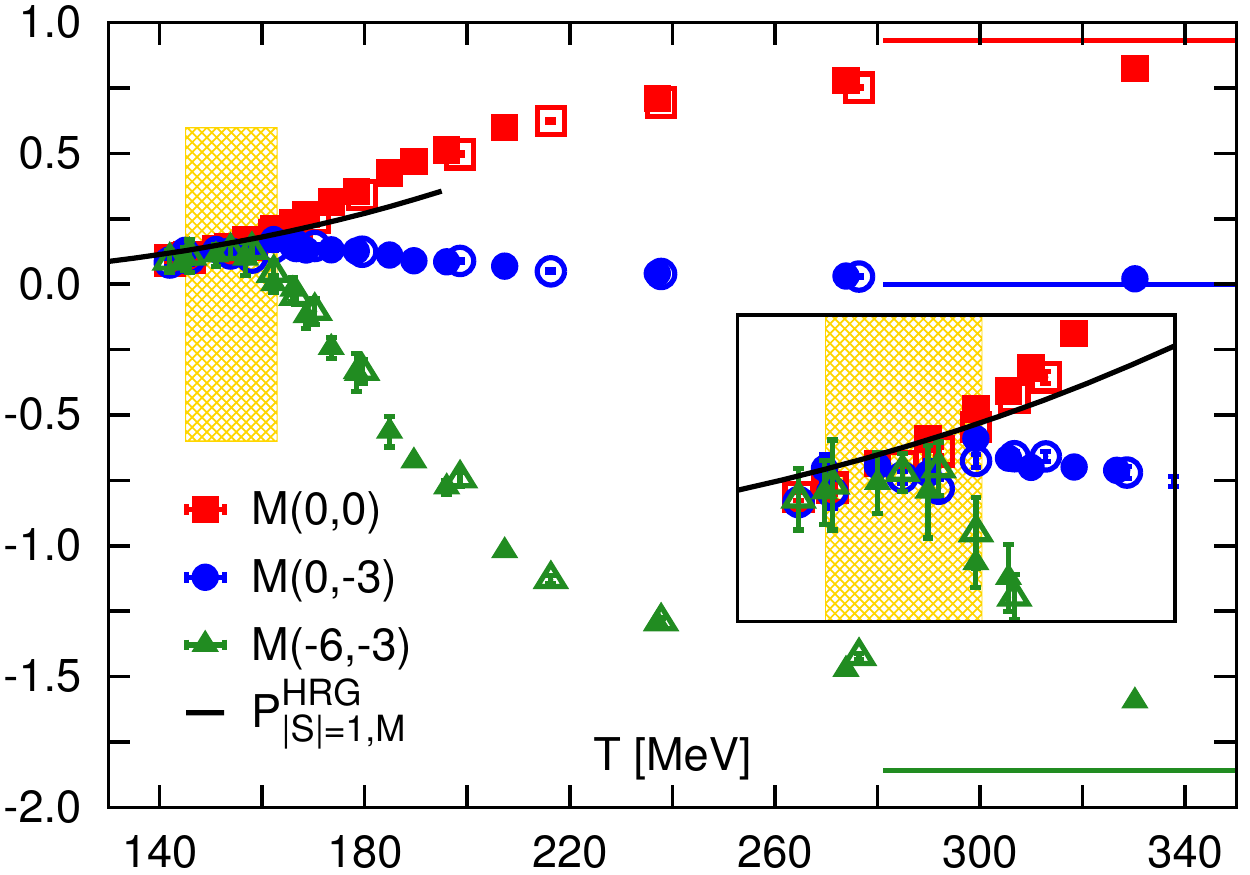}
\end{minipage}\hspace{1pc}%
\begin{minipage}{14pc}
\includegraphics[width=14pc]{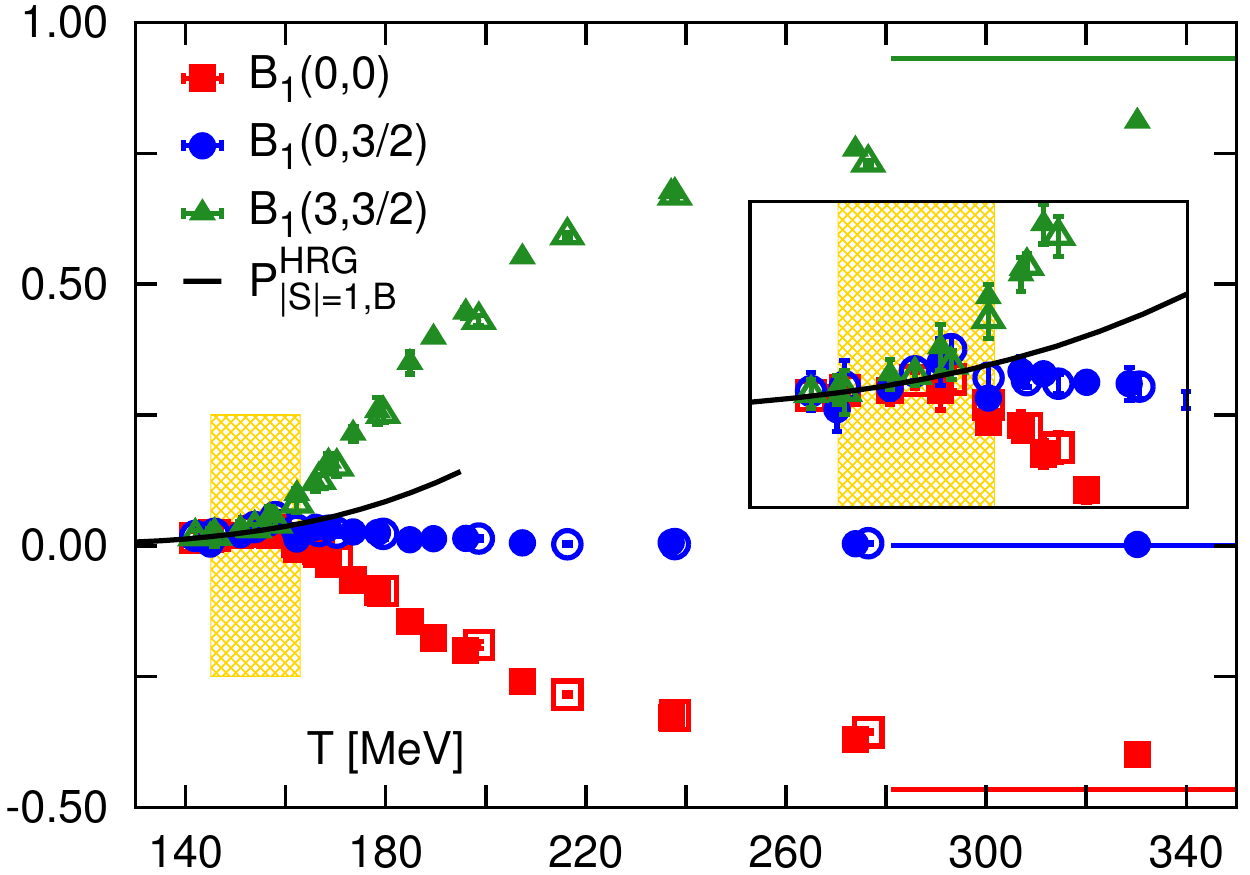}
\end{minipage} 
\begin{minipage}{14pc}
\includegraphics[width=14pc]{M.pdf}
\end{minipage}\hspace{1pc}%
\begin{minipage}{14pc}
\includegraphics[width=14pc]{B1.pdf}
\end{minipage} 
\end{center}
\caption{Operators that project for low temperatures onto the partial pressure of strange mesons (top-left), 
baryons with strangeness $|S|=1$ (top-right), $|S|=2$ (bottom-left) and $|S|=3$ (bottom-right). Results from 
$N_\tau=6$ and $8$ lattices are shown by open and full symbols, respectively. 
Also shown are results from the HRG model (black solid lines). Vertical bands indicate the chiral corossover temperature \cite{hotQCDTc}. \label{fig:pp}}
\end{figure}

\begin{figure}[t]
\begin{center}
\begin{minipage}{14pc}
\includegraphics[width=14pc]{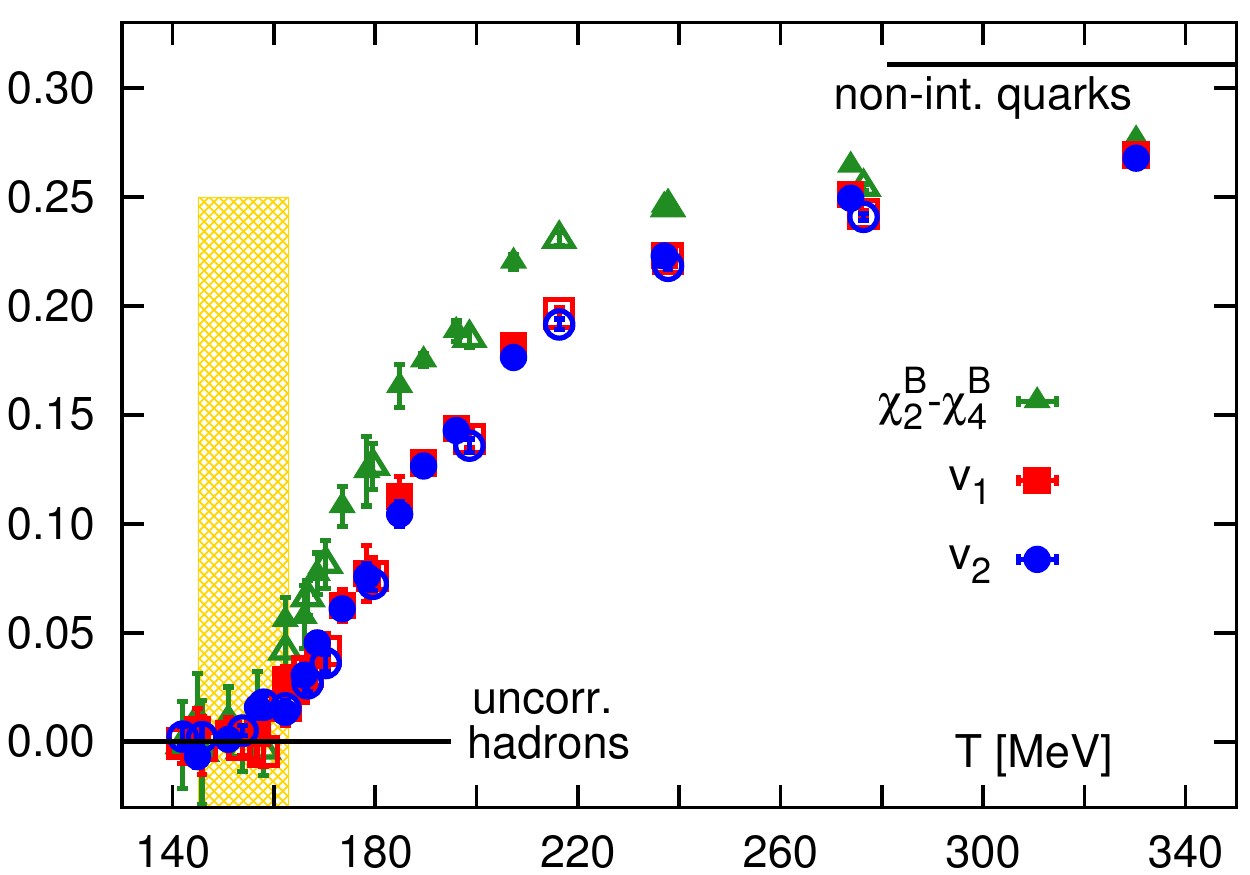}
\end{minipage}\hspace{1pc}%
\begin{minipage}{14pc}
\includegraphics[width=13.4pc]{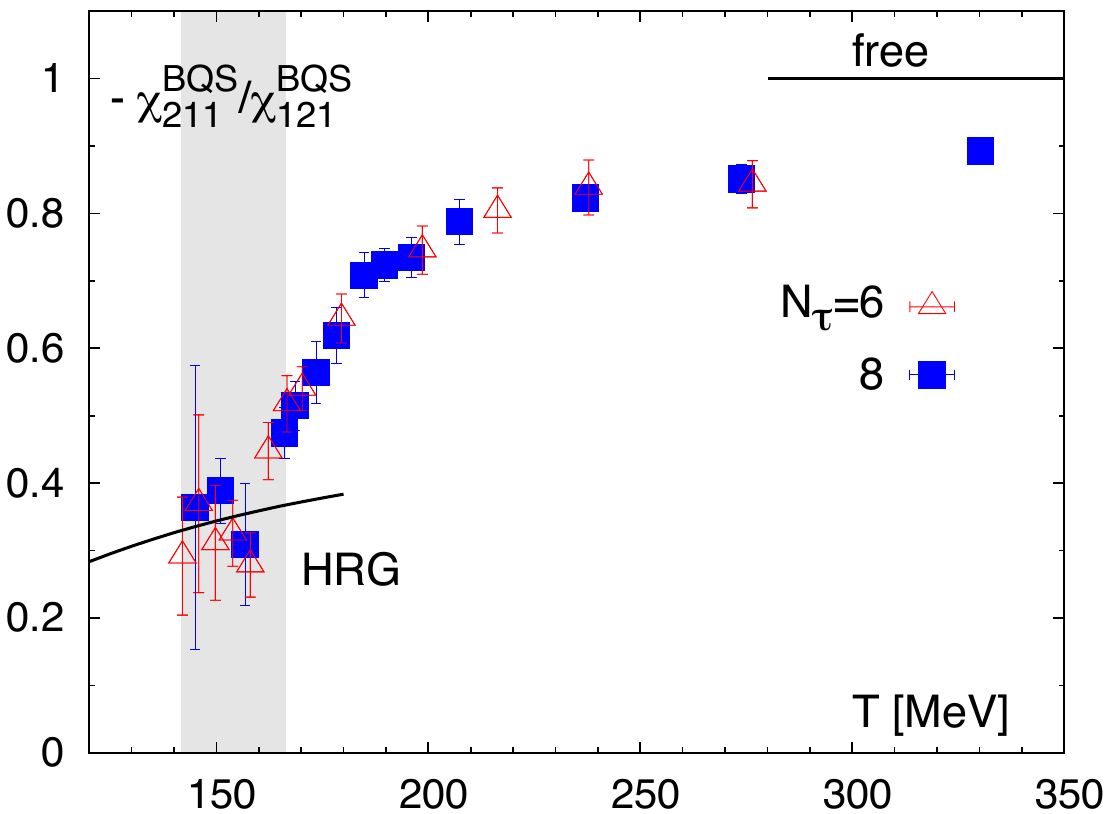}
\end{minipage} 
\end{center}
\caption{Left: quantities that vanish within and uncorrelated gas of hadrons (Boltzmann approximation). Right: quantity that reaches unity in an uncorrelated gas of quark and gluons. In both cases we show results from $N_\tau=6$ and $8$ lattices by open and full symbols, respectively \label{fig:v}}
\end{figure}

Strange hadrons are relatively heavy. Even the lightest strange meson ($K^\pm$) is with $493$ MeV heavy enough to be treated in Boltzmann approximation, in the 
relevant temperature interval just around the chiral crossover ($T_c=154\pm 9$ MeV \cite{hotQCDTc}). The pressure that is build up within an uncorrelated gas of all known 
strange hadrons and resonances $(P^{\rm HRG}_S)$ receives contributions from 4 different strangeness sectors, namely from strange mesons $(|S|=1,M)$ and 
strange baryons $(|S|=i,B)$ with $i=1,2,3$, respectively. Assuming Boltzmann statistics, the $\mu_B$ and $\mu_S$ dependence is simple and solely determined by the 
baryonic and strange charges of the hadrons, {\it i.e.} grouped into the 4 different sectors we find for the partial pressure of all strange hadrons:
\begin{eqnarray}
P^{\rm HRG}_S(T,\hmu_B,\hmu_S) &=& P^{\rm HRG}_{|S|=1,M}(T) \cosh(\hmu_S) 
+ P^{HRG}_{|S|=1,B}(T) \cosh(\hmu_B-\hmu_S)
\nonumber \\
&+& P^{HRG}_{|S|=2,B}(T) \cosh(\hmu_B-2\hmu_S)
+ P^{HRG}_{|S|=3,B}(T) \cosh(\hmu_B-3\hmu_S) \;.
\label{eq:p}
\end{eqnarray}
As long as strange hadrons, with baryonic and strange charges that fall into the before mentioned strangeness sectors, are a valid description of the strangeness 
carrying degrees of freedom in the system, Eq.~(\ref{eq:p}) will imply various relations between correlations of net baryon number and net strangeness $\chi_{mn}^{BS}$, that can be checked in (lattice) QCD. In particular we can use these relations to calculate the partial pressures of the different strangeness sectors from all fluctuations and correlations $\chi_{mn}^{BS}$ up to fourth order, {\it i.e.} with $0\le m\le 4$, $1\le n \le 4$, $(m+n)\le4$ and $m+n$ even, we can use 6 different $BS$-correlations. It is easy to see that according to Eq.~(\ref{eq:p}), all 6 $\chi_{mn}^{BS}$ that we use, can be written as linear combinations of the 4 partial pressures. To obtain the partial pressures we have to invert this linear mapping. We find that the linear mapping has a kernel of dimension 2, consequently our solution is not uniquely defined but will depend on two parameters $c_1$ and $c_2$. As a result, we find the operators that project onto the 4 different partial pressures, given by 
\begin{eqnarray}
M(c_1,c_2) &=& \chi_2^S -\chi_{22}^{BS} + c_1 v_1 + c_2 v_2 \;, \label{eq:M} \\
B_1(c_1,c_2) &=&  \frac{1}{2} \left( \chi_4^S - \chi_2^S +5 \chi_{13}^{BS}+
7 \chi_{22}^{BS} \right) + c_1 v_1 + c_2 v_2   \;, \label{eq:B1} \\
B_2(c_1,c_2) &=& - \frac{1}{4} \left( \chi_4^S - \chi_2^S + 4 \chi_{13}^{BS} +
4 \chi_{22}^{BS} \right) + c_1 v_1 + c_2 v_2 \;, \label{eq:B2}\\
B_3(c_1,c_2) &=& \frac{1}{18} \left( \chi_4^S -  \chi_2^S + 3 \chi_{13}^{BS}+
3 \chi_{22}^{BS} \right) + c_1 v_1 + c_2 v_2\;, \label{eq:B3} 
\end{eqnarray}
where $v_1 = \chi_{31}^{BS} - \chi_{11}^{BS}$ and $v_2 = \frac{1}{3} (\chi_2^S - \chi_4^S ) - 2 \chi_{13}^{BS} - 
4 \chi_{22}^{BS} - 2 \chi_{31}^{BS}$ are 2 linear independent basis vectors of the kernel. Therefore, as long as the uncorrelated gas of hadrons is a valid description of QCD, the combinations $v_1$ and $v_2$ have to vanish\footnote{Of course, the uncorrelated gas model will be an approximation at all temperatures and $v_1$ and $v_2$ are thus not be expected to vanish exactly , {\it i.e.} they are no order parameters. } and do not contribute to Eqs.~(\ref{eq:M}-\ref{eq:B3}).
In Fig.~\ref{fig:pp}, we show our results for the partial pressure, obtained in the 4 different strangeness sectors,
{\it i.e.} we plot our data for the strangeness fluctuations and correlations in the combinations given by Eqs.~(\ref{eq:M}-\ref{eq:B3}) for 3 different choices of the parameter $c_1$ and $c_2$, respectively (color coded).  We find that for temperatures $T\lesssim 160$ MeV, all three different combinations agree, in each of the four sectors. In addition we find agreement with the HRG results shown as solid black lines, in the same temperature range. One can thus conclude that for $T\lesssim 160$ MeV, an uncorrelated gas of hadrons is compatible with the relevant strange degrees of freedom in QCD. The same conclusion can be drawn from Fig.~\ref{fig:v} (left), where we show the combinations $v_1$ and $v_2$, that serve as an indicator for the validity of the HRG,
{\it i.e.} they vanish as long as the strange degrees of freedom can be described by hadrons. Here we also show the difference of second and fourth order net baryon number fluctuations ($\chi_2^B-\chi_4^B$). This quantity also vanishes in the HRG but, in contrast to $v_1,v_2$ receives contributions from all baryons, including the light sector. All quantities shown in Fig.~\ref{fig:v} (left) behave rather similar and in particular break away from the HRG in the crossover region $T\sim (154\pm9)$ MeV. We thus have no indication that strange hadrons survive the chiral crossover. 

\section{The plasma phase}
Finally we probe the strange degrees of freedom in the high temperature phase, were we expect to find strangeness to be carried by quasi free quarks. As strange quarks carry baryon number $1/3$ and electric charge $-1/3$, an interchange of derivatives with respect to $\hmu_B$ and $\hmu_Q$ in a gas of free strange quarks will only lead to a sign change. The ratio $-\chi_{211}^{BQS}/\chi_{121}^{BQS}$ thus reaches unity for large temperatures were quarks behave as free particles. 
In Fig.~\ref{fig:v} (right), where we plot this ratio, we find deviations from the free quark gas result, which are of the order of $(10-15)\%$ for $T\gtrsim 300$ MeV. Similar deviations have also been found in other bulk thermodynamic quantities like pressure and energy density. In the hadronic sector, this ratio projects onto strange charged baryons. Here the charge one and two sectors are weighted differently by the cumulants $\chi_{211}^{BQS}$ and $\chi_{121}^{BQS}$ and the
ratio will be less than unity. We have also plotted the HRG result in Fig.~\ref{fig:v} (right), and find agreement with the HRG for $T\lesssim 160$ MeV.

\section*{Acknowledgement}
CS acknowledges support by the BMBF under grant 05P12PBCTA, the numerical calculations have been performed using the USQCD
GPU-clusters at JLab, the Bielefeld GPU cluster and the NYBlue at the NYCCS.


\section*{References}
\begin{thebibliography}{9}
\bibitem{hic-rev} 
For recent reviews see: 
B. V. Jacak and B. Muller, Science {\bf 337}, 310 (2012); 
B. Jacak and P. Steinberg, Phys.\ Today {\bf 63N5}, 39 (2010).

\bibitem{strange-expt}
C. Blume and C. Markert, Prog.\ Part.\ Nucl.\ Phys.\  {\bf 66}, 834 (2011);
G. Agakishiev {\it et al.}  [STAR Collaboration], Phys.\ Rev.\ Lett.\  {\bf 108},
072301 (2012);
B. I. Abelev {\it et al.}  [STAR Collaboration], Phys.\ Rev.\ C {\bf 81}, 044902
(2010);
M. Nicassio [ALICE Collaboration], Acta Phys.\ Polon.\ Supp.\  {\bf 5}, 237 (2012);
C. E. P. Lara [ALICE Collaboration], arXiv:1303.6496 [nucl-ex].


\bibitem{WHRG}
S. Borsanyi {\it et. al} JHEP {\bf 1201}, 138 (2012).

\bibitem{hotQCDHRG}
A. Bazavov {\it et al.}  [HotQCD Collaboration], Phys.\ Rev.\ D {\bf 86}, 034509
(2012).

\bibitem{WB-Tc}
Y. Aoki, Z. Fodor, S. D. Katz and K. K. Szabo, Phys.\ Lett.\ B {\bf 643}, 46 (2006);
Y. Aoki {\it et al.}, ÊJHEP {\bf 0906}, 088 (2009);
S. Borsanyi {\it et al.}  [Wuppertal-Budapest Collaboration], JHEP {\bf 1009}, 073
(2010);

\bibitem{Ratti} 
C. Ratti, R. Bellwied, M. Cristoforetti and M. Barbaro, Phys.\ Rev.\ D {\bf 85},
014004 (2012);  R.~Bellwied, S.~Borsanyi, Z.~Fodor, S.~DKatz and C.~Ratti, 
arXiv:1305.6297 [hep-lat].

\bibitem{our} 
  A.~Bazavov, H.~-T.~Ding, P.~Hegde, O.~Kaczmarek, F.~Karsch, E.~Laermann, Y.~Maezawa and S.~Mukherjee {\it et al.},
 Phys.\  Rev.\  Lett.\  111, {\bf 082301} (2013).
 
 \bibitem{Koch}
V. Koch, A. Majumder and J. Randrup, Phys.\ Rev.\ Lett.\  {\bf 95}, 182301 (2005).

\bibitem{Ejiri}
S. Ejiri, F. Karsch and K. Redlich, Phys.\ Lett.\ B {\bf 633}, 275 (2006).

\bibitem{GG} 
  S.~Datta, R.~V.~Gavai and S.~Gupta,
  Nucl.\ Phys.\ A904-905 {\bf 2013}, 883c (2013)
  [arXiv:1210.6784 [hep-lat]].

\bibitem{CS} 
  C.~Schmidt,
   Prog.\ Theor.\ Phys.\ Suppl.\  {\bf 186}, 563 (2010)
  [arXiv:1007.5164 [hep-lat]].
  
  \bibitem{freeze-out} 
  A.~Bazavov, H.~T.~Ding, P.~Hegde, O.~Kaczmarek, F.~Karsch, E.~Laermann, S.~Mukherjee and P.~Petreczky {\it et al.},
   Phys.\ Rev.\ Lett.\  {\bf 109}, 192302 (2012)
  [arXiv:1208.1220 [hep-lat]].
  
\bibitem{hotQCDTc}
A. Bazavov {\it et al.}, Phys.\ Rev.\ D {\bf 85}, 054503 (2012).


 \end{thebibliography}
\end{document}